\documentclass[pre,twocolumn,superscriptaddress,showpacs,preprintnumbers,amsmath,amssymb,floatfix]{revtex4}

\usepackage{epsfig,graphicx}
 \usepackage[normalem]{ulem}
 \usepackage{color}
\usepackage{bm}
\def\s#1{_{\rm #1} }
\def\sp#1{^{\rm #1} }
\def\Tr#1{{\rm Tr}\left( #1 \right)}
\def\Det#1{{\rm Det}\left( #1 \right)}

\def\n{{\hat{\bf{n} }}}
\def\k{{\hat{\bf{k} }}}
\def\c{{\hat{\bf{c} }}}
\def\p{{\hat{\bf{p} }}}
\def\y{{\hat{\bf{y} }}}
\def\x{{\hat{\bf{x} }}}
\def\z{{\hat{\bf{z} }}}

\def\R{{\bf{R} }}

\def \be{\begin{equation}}
\def \bea{\begin{eqnarray}}
\def \ee{\end{equation}}
\def \eea{\end{eqnarray}}

\def\half{{\textstyle \frac{1}{2}}}

\newcommand{\bdm}{\begin{displaymath}}
\newcommand{\edm}{\end{displaymath}}

\newcommand{\ten}[1]{\uuline{#1}{}}     

\renewcommand{\vec}[1]{{\mathbf #1}}        

\def\bea{\begin{eqnarray}}
\def\eea{\end{eqnarray}}
\def \be{\begin{equation}}
\def \ee{\end{equation}}

\newcommand{\xibold}{\mbox{\boldmath{$\xi$}}}

\begin{document}

\title{Mechanical switching of ferro-electric rubber}
\author{J.~M. Adams and  M. Warner}
\affiliation{Cavendish Laboratory, University of Cambridge, Madingley Road, Cambridge CB3 0HE, U.K.}
\date{\today}

\begin{abstract}
  At the A to C transition, smectic elastomers have recently been
  observed to undergo $\sim$35\% spontaneous shear strains. We first
  explicitly describe how strains of up to twice this value could be
  mechanically or electrically induced in Sm-$C$ elastomers by rotation
  of the director on a cone around the layer normal at various elastic
  costs depending on constraints.  Secondly, for typical sample
  geometries, we give the various microstructures in Sm-$C$ akin to those
  seen in nematic elastomers under distortions with constraints. It is
  possible to give explicit results for the nature of the textures.
  Chiral Sm-$C$ elastomers are ferro-electric. We calculate how the
  polarization could be mechanically reversed by large, hard or soft
  strains of the rubber, depending upon sample geometry.
\end{abstract}
 \maketitle

\section{Introduction}
One of the most remarkable properties of liquid crystals is their
(improper) ferro-electricity when in the chiral smectic C (Sm-$C$*)
phase \cite{Meyer:75,Lagerwallbook}.  Unlike crystalline
ferro-electrics, their polarization is easily switched, for they are
liquids, albeit with orientational (nematic) and layered (smectic)
order. Thus new types of displays have been based on the fast and
ready response of such liquids to electric fields
\cite{Clark:80,Lagerwallbook,Ljubljanabook,Lagerwall:04}.  We model
the mechanically-driven switching of soft-solid analogues -- Sm-$C$*
elastomers with both orientation and layering where imposed strains
can rotate the polarization, $\vec{P}$.

\begin{figure}[!b]
\begin{center}
\includegraphics[width=0.50\textwidth]{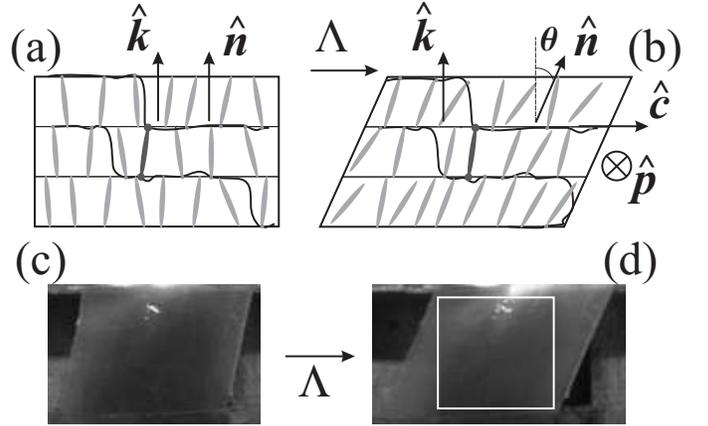}
\end{center}
\caption{(a) Smectic A phase with parallel director $\n$ and layer normal $\k$.
  The polymer backbones are for the polymeric case with the rods shown
  as pendant.  A crosslink (dots at either end of the rod) links
  polymers into the network. (b) Smectic C phase with $\n$ tilted by
$\theta$ in a  direction defined by the unit vector $\c$ in the
plane.  The third (unit) direction, $\p = \k \wedge \c$, is into the
page. It defines the direction of polarization, $\vec{P} = P \p$.
For Sm-$C$ elastomers, there is a spontaneous shear $\Lambda$ with
respect to its Sm-$A$ parent. (c) and (d) Photographs \cite{hiraoka:05}
of the A to C transition in elastomers.  The cut out in (d) is the
Sm-$C$ sample in Fig.~\ref{fig:cartoon_C}(a) that is to be deformed in
this paper.} \label{fig:cartoon}
\end{figure}

Nematic order, about an ordering director $\n$, is along the layer
normal $\k$, in smectic A (Sm-$A$) phases.  RB Meyer {\it et al}
\cite{Meyer:75} recognised that when $\n$ tilts with respect to $\k$
on entering the Sm-$C$ phase, then in a chiral system polarization can
exist: denote the direction of the projection of $\n$ in-plane by
the unit vector $\c$, whence $\k \wedge \c \rightarrow \p $ is an
operation that defines a polar unit vector $\p$ in a chiral system.
Fig~\ref{fig:cartoon} shows the A and C phases, with smectically
ordered rods rather more sharply confined to layers than is
realistic.
The spontaneous polarization is a consequence of ferro-electricity
that is  termed ``improper'' since the order parameter is related to
the tilt \cite{Lagerwallbook} rather than the polarization itself as
in the case of normal ferro-electrics.

Liquid crystalline (LC) polymers display the same phases as
classical materials.  The rods, in the sketch fig~\ref{fig:cartoon},
are pendant to main chains which can then be linked to form an LC
elastomer.  Nematic elastomers suffer large mechanical
elongations/contractions on cooling/heating to and from the ordered
state.  If strains are applied non-coaxially with $\n$, director
rotation and further sympathetic shears can develop to allow shape
change without energy cost in ideal systems, and with little energy
cost in non-ideal systems --- so-called soft elasticity
\cite{kundler:95,warnerbook:07}.  The magnitude of spontaneous
distortion on entering the nematic state sets the scale for the
extent of soft deformation when mechanically-induced director
rotation occurs.

Sm-$A$ elastomers are not soft because (a) the director is not free to rotate without taking the layers with it and
(b) the matrix can only deform while affinely convecting its embedded layers with it and respecting the constancy
of layer spacing. The smectic layer modulus is much larger than the rubber modulus and hence distortions such as
extension along $\k$ are very expensive and in most systems only occur to small amplitude
\cite{Nishikawa:99,adams:05} before instabilities arise. Essentially Sm-$A$ rubbers behave 2-dimensionally; they
stretch and contract in-plane only.  Their shears have either displacements purely in-plane or, if out of plane,
they act to rotate the layers. Sm-$A$ rubber elasticity is highly complex and non-linear.

\begin{figure}[!h]
\begin{center}
\includegraphics[width=0.50\textwidth]{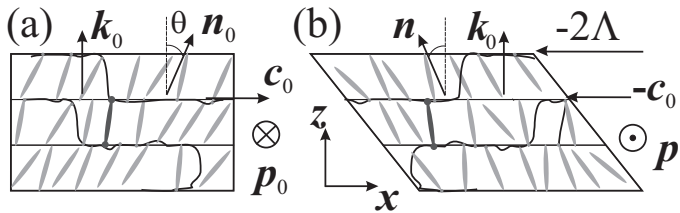}
\end{center}
\caption{A block of Sm-$C$ rubber (a) initially undistorted with $\n_0$
  and $\c_0 \equiv \x$. The layer normal will be taken to remain along
  $\z$. (b) sheared by $\lambda_{xz} = -2 \Lambda$ with respect to its
  relaxed state. The in-plane director has reversed, $\c \rightarrow
  -\c_0$ and thus also the polarization direction: $\p_0 \rightarrow \p = -\p_0$.} \label{fig:cartoon_C}
\end{figure}

The same constraints of constancy of layer spacing act on Sm-$C$
elastomers. The transition Sm-$A$$\rightarrow$Sm-$C$ is accompanied by a
spontaneous shear, $\Lambda$, not trivially related to the molecular
tilt angle $\theta$
\cite{hiraoka:05,Stenull:05,adams:06,Lubensky:06}, as well as an
elongation in plane and contraction along the layer normal. The
spontaneous  shear has been shown to be large, $\Lambda \sim
0.3-0.4$ in the experiments of \cite{hiraoka:05},
Fig.~\ref{fig:cartoon}(c) and (d). We take subsequent distortions
with respect to the spontaneously distorted shape of
Fig.~\ref{fig:cartoon}(b) or (d), that is the reference state shown
relaxed and without distortion in Fig.~\ref{fig:cartoon_C}(a).
  (A residual shear visible in the Sm-$A$ state, Fig.~\ref{fig:cartoon}(c), is an artefact of the 2-step
cross-linking method of achieving monodomains.)

Sm-$C$ elastomers can theoretically be soft since the vector $\vec{n}$
and hence $\c$ can rotate about $\k$ and in doing so induces shape
changes of the body without change of the smectic or rubber elastic
energy. Soft elasticity has been recovered and in general is
predicted to be of considerable complexity in Sm-$C$ elastomers since
the layer normal can also rotate \cite{adams:05C,Stenull:05}.  One
can find \cite{adams:05C} concrete examples of soft modes where the
layer normal remains fixed. We denote the angle of rotation of $\c$
about an unchanging layer normal $\k_0$ by $\phi$. The soft shape
changes associated with changing $\phi$ are shears which conclude at
$\phi = \pm \pi$, with $\lambda_{xz} = -2\Lambda$ and all other
distortions vanishing. In Fig.~\ref{fig:cartoon_C}(b) this final
shear corresponds to reversing the spontaneously sheared shape in
Fig.~\ref{fig:cartoon}(b) to its opposite form.  The spontaneous
simple shear thus has an important role in delineating the extent of
softness of imposed deformations in smectic rubbers. Analogously, in
experiment and theory of softness in nematic elastomers, the extent
is delineated by the extent of spontaneous elongation on entering
the nematic phase. For $\phi$ increasing still further, the original
undistorted state is eventually regained at $\phi = 2\pi$.

As $\c$ rotates about $\k$ by $\phi$, then so too does the
polarization direction $\p$.  If it is initially along $\y$, then
when $\c$ has rotated by $\pi$, $\p$ has reversed to $-\y$, see
Fig.~\ref{fig:cartoon_C}(b).   Much of this paper is concerned with
describing how this reversal of polarization can be achieved by the
imposition of shear deformation to the elastomer. The applications
of shear generating large electrical changes are obvious and very
attractive.
 With the rotations of $\vec{c}$ now
defined, we can now give a concrete form of the true order parameter
which  is an in-plane vector $\xibold =
[(\vec{k}\wedge\vec{n})\wedge\vec{k}]\wedge\vec{n}= -
(\vec{k}\wedge\vec{n})(\vec{k}\cdot\vec{n}) = \sin\theta\cos\theta
(-\sin\phi,\cos\phi)$, see \cite{Pikinbook}, and also
\cite{Lagerwallbook} using an analogous form from superconductors.

Although it is possible to find soft trajectories of deformation that
reverse $\p$, these will not in general satisfy boundary conditions
imposed by, say, rigid electrodes or clamps.  One can, in some cases,
take combinations of $\pm \phi$ deformations that form a texture that
overall satisfies the external constraints, that is the free energy
has been ``quasi-convexified'' \cite{Bhattacharyabook:03}, a process
well-understood in the routes to soft deformation of nematic
elastomers \cite{Conti:02,warnerbook:07}. We show different textures
for Sm-$C$ elastomers that depend on the constraints offered by the
two sample geometries.  The general mathematical problem of how
textures in Sm-$C$ elastomers make possible soft deformations in the
presence of constraints has been attacked by Adams {\it et al}
\cite{adams:06a}.  In particular it is possible to find geometries
that are soft under tension and hence easier to verify experimentally.
Otherwise, Sm-$C$ elastomers can deform via non-soft alternatives that
fully satisfy constraints. These will also require the development of
textures during deformation, and we calculate some of them here. In
either case there will be barriers (possibly smaller for soft
textures) between the two states of reversed polarization. The choice
between soft and non-soft alternatives will depend upon whether one
has sheet or slab geometry -- we discuss both choices in
Section~\ref{sect:sample}.  In a companion paper to this
\cite{Biggins_Bhattacharya:09} it is shown that there are two possible
types of stripe-domain, or texture, of which our textures that do not
involve the rotation of smectic layers across the laminate boundary
are drawn from one class.  For chiral Sm-$C$ elastomers, the internal
boundaries are in general charged for the class we are dealing with;
the other class has uncharged internal boundaries
\cite{Biggins_Bhattacharya:09}.

 Most experiments on Sm-$A$ \& C  elastomers
have involved mechanical and electrical changes to the tilt angle,
e.g. \cite{lehmann:01}.  We are looking at Goldstone modes that
instead rotate the director about the layer normal at essentially
fixed cone angle of tilt.  Experiments on polydomain Sm-$C$ elastomers
\cite{Sanchez:08} show that this rotation is easily possible and can
in fact lead to gross reorganization of the domain structure, even
its removal. Being polydomain the deformation was not especially
soft, but demonstrated LC mobility in this phase and offers hope for
the deformations we predict here.  In any event, we hope that the
range of deformation paths we describe will urgently stimulate
experiment to explore the ferro-electric response of smectic
elastomers to shears opposed to the spontaneous distortion that
arises on leaving the A-state. Our theoretical models will show how
completely open the understanding of these systems is.

\section{A model for non-linear distortions of smectic-C
rubber}\label{sect:model} We adopt a particular model that
successfully describes experiments on the non-linear rubbery and
smectic elasticity of Sm-$A$ elastomers, and which has also been
applied to Sm-$C$ elastomers.  The underlying nematic rubbery
elasticity is subject only to layers moving affinely with the bulk
and then only allowing distortions of the bulk that then respect the
constancy of layer spacing. Thus material points $\R_0$ and layer
normals $\k_0$ in the reference state transform as \cite{adams:05}:
 \be \R = \ten{\lambda}\cdot \R_0 \;\;\; {\rm and}
\;\;\; \k = \ten{\lambda}^{-T}\cdot \k_0 .\label{eq:transforms} \ee
The deformation gradient $\ten{\lambda}_s$ must respect rigid
constraints of constancy of volume and of smectic layer spacing.
This is because the  shear modulus of the isotropic state of the
elastomer and for the smectic for deformations not involving layer
spacing changes is $\mu \sim 10^5-10^6{\rm J/m^3}$, nearly two
orders of magnitude smaller than the smectic layer spacing modulus
for elastomers and 4 orders of magnitude smaller than the bulk
modulus, and hence deformations avoid layer spacing and volume
changes. We discuss in Appendix A how $\mu$ can be estimated by
measurements in the Sm state. Thus $\ten{\lambda}$ is rigidly
constrained so that $|\ten{\lambda}^{-T} \cdot \k_0| = 1$ where $-T$
denotes transposed inverse. $\Det{\ten{\lambda}} = 1$ expresses
constancy of volume.

The free energy density, $f$, and a general soft deformation,
$\ten{\lambda}_s$, are respectively:
\be f = \half \mu \Tr{\ten{\ell}_0\cdot \ten{\lambda}^T \cdot
  \ten{\ell}_n^{-1} \cdot \ten{\lambda}} ; \;\;
\ten{\lambda}_s = \ten{\ell}_n^{1/2} \cdot \ten{W} \cdot
\ten{\ell}_0^{-1/2} \label{eq:energy}
\ee
with $\ten{W}$ a general rotation matrix in the nematic case
\cite{olmsted:94}, and a more specialised form for smectic
elastomers \cite{adams:05C}. The shape tensors of the network
polymers, $\ten{\ell}_0$ initially and $\ten{\ell}_n$ after
deformation (when the director $\n_0$ has possibly rotated to the
new direction $\n$) are of the form $\ten{\ell}_0 = (r-1) \n_0 \n_0
+ \ten{\delta}$ (and analogously for $\ten{\ell}_n$) where $r$ is
the anisotropy in the distribution of polymer shapes. These tensors
encode information about the direction of order. We take the
distribution for simplicity to be uniaxial. This approximation is
discussed in \cite{adams:05C} --- effects due to rotation of the
long axis of the anisotropic distribution are very large and give a
first order description of a system that in reality is certainly to
an extent biaxial. The minimum free energy is $3\mu/2$ which obtains
when there is no distortion, $\ten{\lambda} = \ten{\delta}$ and no
director rotation, $\n = \n_0$ and hence $\ten{\ell}_n^{-1} =
\ten{\ell}_0^{-1}$, or when deformations are soft, $\ten{\lambda}^s
= \ten{\lambda}$.

The spontaneous shear $\lambda_{xz}\sp{c}$ in this model
\cite{adams:06}, denoted here by $\Lambda$, is given by
\be
\lambda_{xz}\sp{c} \equiv \Lambda = (r-1)\sin\theta\cos\theta/\rho
\label{eq:spont}
\ee
where the combination $\rho = r\cos^2\theta + \sin^2\theta \le r$
will repeatedly follow in the concrete examples of distortions we
shall give.  The molecular details are thus simply encoded.  At
fixed temperature, $\theta$ does not change if we also assume that
anchoring is rigid, that is applied strains do not mechanically
alter the tilt angle.  In considering elastomers with extreme
anisotropy of mechanical properties, we are implicitly dealing with
smectic elastomers where the layer structure and other details of
molecular ordering are on a higher energy scale than rubber
elasticity \cite{Nishikawa:99}. There may be systems in which tilt
is not rigid under imposed strains, a possibility that has been
considered theoretically \cite{Stenull:05,Lubensky:06,Stenull:08}
and experimentally \cite{Kramer:08}. It is possible that the chain
anisotropy $r$ might change with tilt $\theta$ and thus that the
$\theta$ dependence is more complicated than appears in
eqn~(\ref{eq:spont}). This complication will not concern us for
elastomers at fixed temperature, and thus fixed tilt, during
mechanical experiments.  We are not dealing with smectic elastomers
in which layers do not appear to significantly effect mechanical
properties, for instance elastomers where one can induce compression
of the smectic layers by applying an in-plane strain
\cite{Stannarius:06}.  Likewise we are not addressing experiments
that (i) use mechanical compression to alter the tilt angle (thereby
changing $P$ and hence a route to piezoelectricity) or (ii) apply an
electric field, alter tilt and hence generate strain along the layer
normal (an inverse piezoelectric effect)
\cite{lehmann:01,Koehler:05,Spillmann:07}.

We now consider 3 explicit types of deformation in response to
imposed shears that seek to redirect  the spontaneous shears
observed on the A to C transition.  The deformations are treated
separately.  They have increasing freedom to exercise various
sympathetic distortions which serve to progressively soften the
elastic energy penalty.

\subsection{Simple, non-soft response to shear}
We are interested in shearing Sm-$C$ elastomers simply, without rotating
or distorting the layers, perhaps by fixing rigid plates to their $xy$
surfaces. Firstly, consider a simple, non-soft deformation with its
inverse transpose:
\bea \ten{\lambda} = \left(
\begin{array}{ccc}
1&0&\lambda_{xz}\\
0&1&\lambda_{yz}\\
0&0&1
\end{array}
\right); \;\;\; \ten{\lambda}^{-T}=\left(
\begin{array}{ccc}
1&0&0\\
0&1&0\\
-\lambda_{xz}&-\lambda_{yz}&1
\end{array}
\right). \label{eqn:simple_hard}    \eea
Trivially $\Det{\ten{\lambda}} = 1$, volume is conserved, and
$\ten{\lambda}^{-T} \cdot \z = \z$, the layer normals are not rotated
and their separation is unchanged by the action of $\ten{\lambda}$.
The 0 and 1 entries in (\ref{eqn:simple_hard}) ensure that the $xy$
plates do not change shape.

We take an initial $\c_0 = \x$ as in Fig.~\ref{fig:cartoon_C}(b) and
impose a shear $\lambda_{xz}$.  The shear $\lambda_{yz}$ is the
relaxation expected as $\n$ is induced to rotate by $\phi$ about
$\k$ towards $\y$.  Inserting $\ten{\lambda}$ into $f$, and
minimizing over $\lambda_{yz}$ and $\phi$ gives the optimal free
energy and relaxation as a function of $\lambda_{xz}$.  The
$\ten{\lambda}$ of eqn~(\ref{eqn:simple_hard}) is not soft: since
the director is being rotated, the elongation associated with it is
also rotated so one expects extension along $y$, contraction along
$x$ and some $yx$ shear in plane (see below where we successively
allow these relaxations).  There is a cost of constraining diagonal
elements to 1 and some shears to 0, leading to a threshold,
$\lambda_1$, before $\lambda_{yz}$ relaxation and rotation of $\c$
starts.  Until then the free energy is hard with a corresponding
modulus, see the curvature of the initial part of
Fig.~\ref{fig:hard_deformation}(a), and full details in Appendix A.
For the $r=2$, $\theta = \pi/6$ chosen for illustration throughout,
the spontaneous shear would be $\Lambda =\sqrt{3}/7 \simeq 0.246$.
\begin{figure}[!t]
\begin{center}
\includegraphics[width=0.5\textwidth]{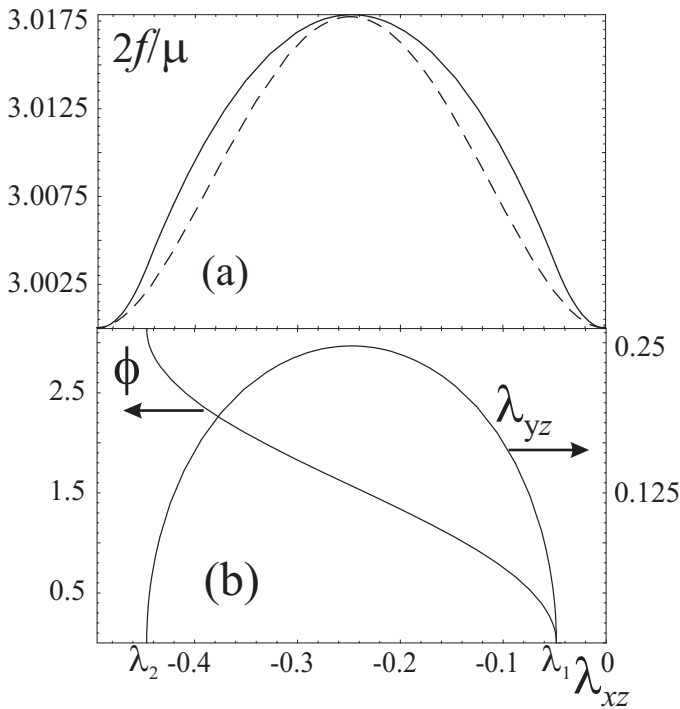}
\end{center}
\caption{(a) The elastic free energy, in units of $\half \mu$, against
  imposed shear $\lambda_{xz}\leq 0$.  The dotted line is energy cost when allowing
  $yx$ relaxation as well. (b) Rotation $\phi$ of the in-plane director $\c$
  about the layer normal, and the concomitant $yz$ shear relaxation,
  both starting and concluding at thresholds $\lambda_1$ and
  $\lambda_2$ respectively. The anisotropy is $r=2$ and the director
  tilt is $\theta = \pi/6$.}
\label{fig:hard_deformation}
\end{figure}
When relaxation starts, the free energy is softened considerably
since now shape change can be more by rotation of the long axis of
the polymer chain distribution than by an expensive distortion of
the chains. The shear $\lambda_{yz}$ and rotational $\phi$
relaxation, Fig.~\ref{fig:hard_deformation}(b), are both initially
singular. Director rotation is directly observable optically.  We
defer discussing it until we consider sample geometry and the
questions of textures and ferro-electric switching.  The thresholds
are slightly complicated functions of $r$ and $\theta$ that can be
found analytically from the free energy.  The singular forms arise
because both $\pm\lambda_{yz}$ and $\pm\phi$ give the same
$\lambda_{xz}$. Thus for instance $\lambda_{xz} \propto \phi^2$ or
conversely there is a square root singularity $\phi \propto
\sqrt{\lambda_{xz}}$, and similarly for $\lambda_{yz}$.

The $yz$ shear reaches a maximum numerically equal to $\Lambda$ at
$\phi = \pi/2$ and $\lambda_{xz} = -\Lambda$ since at this imposed
shear $\c$ points along $\y$. The anisotropy now along $\y$ is
 analogous to the tilt of the anisotropy along $\x$ in
Fig.~\ref{fig:cartoon}(b).  The same $yz$-shear develops as was the
case with $xz$-shear in going from Sm-$A$ to Sm-$C$ in
Fig.~\ref{fig:cartoon}.  But the free energy is instead maximal at
this shear.  It then declines back to its minimal value $3\mu/2$
when the imposed shear is $-2\Lambda$.  The $yz$-relaxation
vanishes, and $\phi$ attains $\pi$, in a singular manner at a
threshold, $\lambda_2$, equivalent to that at small $\lambda_{xz}$,
before the $xz$ shear reaches $-2\Lambda$. In non-ideal systems
(semi-soft elastomers) the free energy would not be at an absolute
minimum at $\phi = \pi$. We discuss semi-soft response below in
considering the role of disproportionation. The singular behaviour
is seen and expected for analogous distortions involving
mechanically-induced director rotation in nematic elastomers, see
the experiments of Finkelmann {\it et al} \cite{Kundler1a}.  The
imposition of shear in a slab geometry (discussed later in
figure~\ref{fig:textures}) is straightforward. It is also possible
in sheets of the form of figures~\ref{fig:cartoon}(c) and (d), up to
a certain maximum shear where wrinkling on-sets.  Even corresponding
Sm-$A$ sheets, which are certainly not soft, have been successfully
sheared to investigate mechanically-induced director tilt into the
C-form (the mechano-clinic effect) \cite{Kramer:08,Stenull:08}.

\subsection{Additional, non-soft shear relaxation}
\begin{figure}[htb!]
\begin{center}
\includegraphics[width=0.50\textwidth]{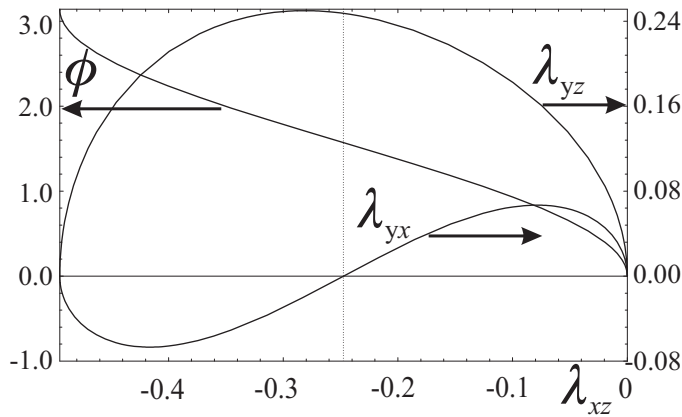}
\end{center}
\caption{The $yx$ and $yz$ shears and in-plane director rotation
$\phi$ against imposed shear $\lambda_{xz}$ for the softer case
where $yx$ relaxation is permitted  ($r$ and $\theta$ as before).  The light vertical dotted line is at $\lambda_{xz} = -\Lambda$ to emphasize the asymmetry about the point where the director is transverse; see Appendix~\ref{app:asymm}.}
\label{fig:softish_deformation}
\end{figure}

It is possible to find slightly softer deformation trajectories than
that illustrated above by including the $yx$ component of deformation.
For systems with clamps allowing changes in the shape of the $xy$
section of the elastomer, or for micro-structures that we will explore
below, this extra freedom might be a way of reducing the elastic
energy between the two minimal states at $\phi=0$ and $\phi = \pi$.
Thus the deformation gradient and its inverse transpose are
respectively
 \bea \left(
\begin{array}{ccc}
1&0&\lambda_{xz}\\
\lambda_{yx}&1&\lambda_{yz}\\
0&0&1
\end{array}
\right)  {\rm and} \;\left(
\begin{array}{ccc}
1&-\lambda_{yx}&0\\
0&1&0\\
-\lambda_{xz}&\lambda_{yx}\lambda_{xz}-\lambda_{yz}&1
\end{array}
\right) \label{eqn:softish}
\eea
where, with the aid of the latter, one can confirm that the layer
normal has not been rotated nor the layer spacing changed by the
imposed deformation.  With this greater freedom, the threshold to
both director rotation and strain relaxation can be avoided.  Thus
the free energy cost is lowered from the previous trajectory, see
the dotted curve also plotted in Fig.~\ref{fig:hard_deformation}(a)
for direct comparison with the energy when less relaxation is
permitted. This minimal free energy arises from  the optimal
deformation gradient $\ten{\lambda}$, with elements explicitly given
below, being put into $f$ of eqn~(\ref{eq:energy}). The same
procedure must be followed still further below when an explicit form
for a soft deformation is given, but the result for $f$ is then
trivial (the unchanging value $3\mu/2$).  Here, the components of
the optimal deformation gradient tensor are:
 \bea
 \lambda_{xz} &=& -\frac{r-1}{\rho} \sin\theta\cos\theta (1-\cos\phi) \label{eq:softish-xz} \\ &\equiv& -\Lambda
 (1-\cos\phi) \nonumber \\
\lambda_{yz} &=& \Lambda    \left(1 - \frac{r-1}{a^2 r} \sin^2\theta \cos\phi\right)\sin\phi
\label{eq:softish-yz} \\
\lambda_{yx} &=& \frac{r-1}{2 a^2 r} \sin^2\theta \sin2\phi\label{eq:softish-yx}
 \eea
 with the combination $a^2 = \cos^2\phi + (\rho/r) \sin^2\phi \le 1$.
 The $yx$ and $yz$ shears plus the accompanying director rotation are
 plotted in Fig.~\ref{fig:softish_deformation} against the imposed
 $xz$ shear.

The rotation is straightforwardly $\cos\phi = 1+ \lambda_{xz}/\Lambda$
from (\ref{eq:softish-xz}) and must be inserted into
eqns~(\ref{eq:softish-yz}) and (\ref{eq:softish-yx}) to get
$\lambda_{yz}(\lambda_{xz})$ and $\lambda_{yx}(\lambda_{xz})$. The
singular rotation initially and finally is just that of the
$\cos^{-1}$ function:
\bea \phi &=& \pm\sqrt{-2\lambda_{xz}/\Lambda} \;\;\;\;  \;\;\;\;{\rm for} \;\; \;\;\;\lambda_{xz}\lesssim 0
 \label{eq:singular0}\\
  \phi &=& \pm\sqrt{2(2-\lambda_{xz}/\Lambda)} \;\; {\rm for} \;\; -2\Lambda \lesssim \lambda_{xz}
 \label{eq:singular2}.
\eea
At $\phi = \pi/2$ one has $\lambda_{xz} = - \Lambda$, the $xz$ section
of the sample has gone half way between the states of
Figs~\ref{fig:cartoon_C}(a) and (b).

The new shear, $yx$, is antisymmetric about $\phi = \pi/2$.  It must
vanish when $\c$ points along $\y$ since the natural change of the
sample would be to elongate along $\y$ (which we as yet suppress,
$\lambda_{yy}=1$) and to $yz$-shear.  There is no need for in-plane
shape change, $\lambda_{yx}=0$.

The $yz$ deformation is numerically equal to $\Lambda$ at $\phi =
\pi/2$ as one would expect -- when the director has rotated by
$\pi/2$ the $y$ axis has a similar significance to that of the $x$
axis initially, but the energy is now high. This  intermediate state
$\phi = \pi/2$ with $\n$ in the $yz$ plane and $\c = \y$, is where
there must be maximal $yz$-distortion due to director rotation
redirecting the elongated dimension of the sample.  Indeed the
energy can be seen to be symmetric about $\pi/2$ (actually seen in
the plots against $\lambda_{xz}$ to be symmetric about the value
$-\Lambda$). However, since we are dealing with large deformations
that do not add linearly, the additional shear introduces an
apparent asymmetry about $\phi = \pi/2$ into the form of the
$yz$-relaxation. (In Appendix~\ref{app:asymm}, we show that this apparent
asymmetry, and more asymmetries to be discussed below, are simply
consequences of the geometric need to compound rather than add large
deformations.) The $yz$ relaxation reaches a numerically maximal
value at $\phi = \half
\cos^{-1}\left[(r-1)\sin^2\theta/\left((r-1)\cos^2\theta
    + r+1 \right)\right]$.

The energy maximum offers the same height of barrier between the
minimal states as before. With respect to the minimal value of
$3\mu/2$, the barrier is $2f\s{bar}/\mu = 2 (r-1)^2
\sin^4\theta/\left[ r(r+(r-1)\cos 2\theta)\right]$. As in all
expressions involving the energy cost on rotation or shear, it
scales as $(r-1)$, that is it vanishes on isotropy $(r=1)$.
Otherwise, the energy is somewhat reduced, see the dotted curve in
Fig~\ref{fig:hard_deformation}(a), but at the expense of a more
complex system of sympathetic shears. We now explore a final shear
scenario to lower the free energy of distortion still further to its
minimal value.

\subsection{Soft deformations}

\begin{figure}[htb!]
\begin{center}
\includegraphics[width=0.50\textwidth]{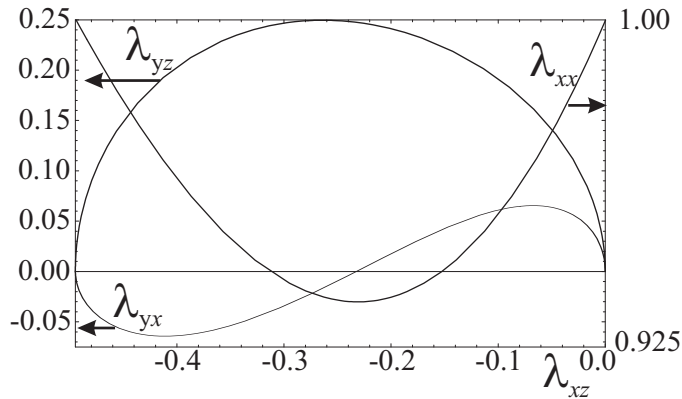}
\end{center}
\caption{
  The $yx$ and $yz$ shears and the $xx$ contraction against imposed
  shear $\lambda_{xz}$ for soft deformations ($r$ and $\theta$ as
  before).}
\label{fig:soft_deformation}
\end{figure}

If the rubber has total freedom to shear and distort as the director rotates, then in ideal systems there is zero
accompanying rubber elastic cost.  The mechanism arises because the distribution of chains is accommodated by the
changing shape of the body without the distribution's distortion and thus with no decrease of the entropy or
modification of the liquid crystal order. Such distortions are well-known in nematic elastomers \cite{kundler:95}
deep into the non-linear regime.  They have been explored theoretically in Sm-$C$ elastomers
\cite{adams:05C,Stenull:05,Lubensky:06} where the constraint of layers must be rigidly observed.  A soft
deformation gradient (and its inverse transpose), which leaves the layer normal unrotated and the layer spacing
unchanged, is
\bea \ten{\lambda} &=& \left(
\begin{array}{ccc}
\lambda_{xx}&0&\lambda_{xz}\\
\lambda_{yx}&\lambda_{yy}&\lambda_{yz}\\
0&0&1
\end{array}
\right) \label{eq:soft}\\ \ten{\lambda}^{-T}  &=& \left(
\begin{array}{ccc}
\lambda_{yy}&-\lambda_{yx}&0\\
0&\lambda_{xx}&0\\
-\lambda_{yy}\lambda_{xz}&\lambda_{yx}\lambda_{xz}-\lambda_{xx}\lambda_{yz}&1
\end{array}
\right) \label{eqn:softT}
\eea
where $\lambda_{zz} = \lambda_{xx}\lambda_{yy} = 1$ from
incompressibility, $\Det{\ten{\lambda}} = 1$.  The elements of the
tensor are:
\bea\lambda_{xx} &=& 1/ \lambda_{yy} = a(\phi) \label{eq:softxx}\\ \lambda_{xz} &=& \Lambda
\textstyle{\left(-a(\phi)+ \cos\phi\right)}\label{eq:softxz}
\\
\lambda_{yx} &=& \frac{r-1}{2 r a} \sin^2\theta \sin 2 \phi
\label{eq:softyx}\\ \lambda_{yz} &=& \Lambda\textstyle{\left(\sin
\phi-\frac{r-1}{2 r a} \sin^2\theta \sin 2
\phi\right)}\label{eq:softyz}, \eea
with $a^2 = \cos^2\phi + (\rho/r) \sin^2\phi$ which was introduced
below eqn~(\ref{eq:softish-yx}), and $\lambda$ given in
eqn~(\ref{eq:spont}). A factor $\left(1-\frac{\rho}{r}\right)$
appearing in the expression derived in \cite{adams:05C} has been
replaced by the equivalent $(r-1)\sin^2\theta / r$.  The same types
of shears as in eqn~(\ref{eqn:softish}) enter, but we allow the
elongations and contractions $\lambda_{xx}$ and $\lambda_{yy}$ to
adjust to the changing natural length in the $x$ and $y$ directions
as the polymer chain distribution anisotropy is rotated by $\phi$.
It is this final element, plus the concomitant further changes to
the shears, that allows the deformation to be soft.
Fig.~\ref{fig:soft_deformation} shows the $xx$ contraction, and the
$yx$ and $yz$ shears against the imposed $xz$ shear.

\begin{figure}[htb!]
\begin{center}
\includegraphics[width=0.50\textwidth]{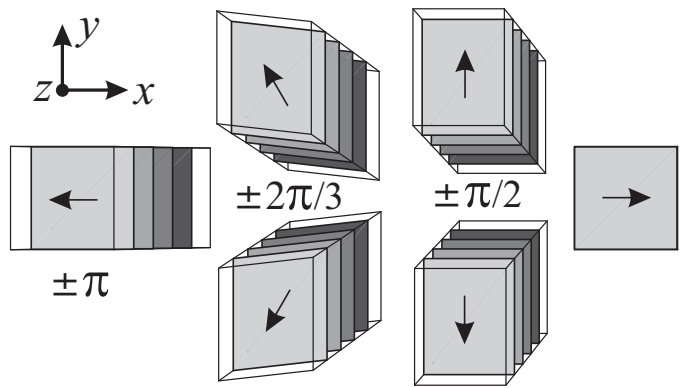}
\end{center}
\caption{
  Soft deformations of a cube of Sm-$C$ rubber with $r=8$ and $\theta =
  \pi/6$ (as before), viewed along the smectic layer normal.  The $\vec{c}$ vectors and the corresponding $\phi$ are shown.}
\label{fig:soft_modes}
\end{figure}

In Fig.~\ref{fig:soft_modes} we show snap shots of an initial cube
deforming under a soft $\ten{\lambda}$ as the $\c$-director advances
(right to left) through $\phi = 0$, $\pm \pi/2$, $\pm 2\pi/3$ and
$\pm \pi$.  It starts from Fig~\ref{fig:cartoon_C}(a) and ends in
(b), but we are viewing it from along the $z$ axis rather than along
the $y$ axis.
Such a monodomain response, either along the $+\phi$ or $-\phi$
route, may not conform to macroscopic constraints, for instance those
that suppress $\lambda_{yz}$ shear.  Much of the remainder of this
paper is about adding $\pm\phi$ distortions in microtextures to
conform to constraints.

The rotation $\phi$ is of the singular form as before, with slight
modifications that are apparent on comparing the soft
$\lambda_{xz}(\phi)$, eqn~(\ref{eq:softxz}), with the corresponding
$xz$ deformation gradient, eqn~(\ref{eq:softish-xz}), in the hard
case with $yx$-relaxation.  Explicitly, $\phi(\lambda_{xz})$ follows
from (\ref{eq:softxz}):
\bea \phi = \cos^{-1}\hspace{-1mm}\left[
\frac{r}{\rho}\hspace{-1mm}\left(
\sqrt{\left(\frac{\rho}{r}\right)^2 +
\left(1-\frac{\rho}{r}\right)\left(\frac{\lambda_{xz}}{\Lambda}\right)^2}
\! +\frac{\lambda_{xz}}{\Lambda} \right) \right]
\label{eq:softphilambdaxz} \eea with singular rotation initially
(and analogously finally): \bea \phi \sim \pm \sqrt{\frac{2r}{\rho
}\lambda_{xz}/\Lambda} \;\;\ {\rm for}\;\; \lambda_{xz} \lesssim 0 .
\label{eq:softphi-limit} \eea
Note that $\lambda_{xz}$ is no longer $-\Lambda$ when the
$\c$-director is along $\y$, that is $\phi = \pi/2$, but is instead
$\lambda_{xz} = - \Lambda \sqrt{\rho/r} \equiv -\Lambda \sqrt{1 -
  \frac{r-1}{r}\sin\theta \cos\theta}$, a numerically smaller value
than before.  Figs~\ref{fig:cartoon_C}(a) and (b) correspond to
shears of $0$ and $-2\Lambda$ and when the system has $\phi = \pi/2$
it is geometrically half way between.  The {\it apparent}
inconsistency because the shear is not $-\Lambda$, is again a
consequence of non-linearity, see Appendix B.  However, when the
$\c$-director is transverse, there is, as expected, no in-plane
shape change, $\lambda_{yx} = 0$, as can be seen in
(\ref{eq:softyx}) for this $\phi$.  The $yz$-shear can be seen to be
$\Lambda$ at $\phi = \pi/2$, but since $\lambda_{xz} > -\Lambda$
there, then $\lambda_{yz}(\lambda_{xz})$ is asymmetric about
$-\Lambda$ as in the previous subsection, that
is because of the geometrical non-linearity of finite deformations.

\section{Sample geometry and textured response}
\label{sect:sample}
\begin{figure}[!htb]
\begin{center}
\includegraphics[width=0.50\textwidth]{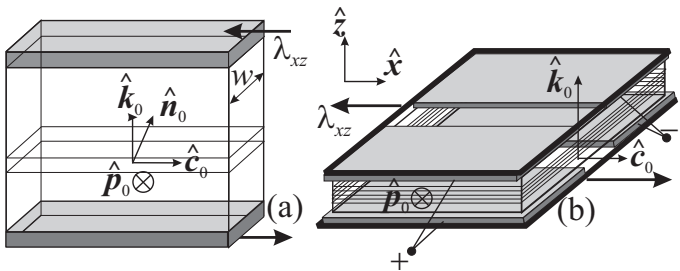}
\end{center}
\caption{
  (a) Sheet and (b) Slab geometries for Sm-$C$ elastomers. Rigid clamps
  (shaded) are shown on the $xy$ sample surfaces. $\lambda_{xz}$ shear
  displacements are indicated by arrows.  A few smectic planes are
  indicated in each case, with normals $\k_0$. polarization $\vec{P}$
  in the $y$ direction (of extent $w$) emerges from the $xz$ sample
  surfaces.  Lines of electric flux in the slab case are captured
  by enveloping the small $xz$ surfaces by overhanging, split $xy$
  electrodes, their charge being identified by $\pm$.}
\label{fig:textures}
\end{figure}

The form of deformation in response to an imposed shear $\lambda_{xz}$
or director rotation $\phi$ (driven perhaps electrically) will depend
on what components of $\ten{\lambda}$ are inhibited by geometrical
constraints.  We now discuss different sample geometries.  Hiraoka
{\it et al} \cite{hiraoka:05} created monodomains and observed large
spontaneous shears on the Sm-$A$ $\rightarrow $ Sm-$C$ transition in a sheet
geometry, Fig.~\ref{fig:textures}(a), where the $y$ dimension of the
sample is small.
Alternatively smectic actuation, albeit from electroclinic effects
in Sm-$A$* elastomers where tilt and shear are purely
electrically-driven, has been achieved \cite{lehmann:01} with small
shears in the slab geometry of Fig.~\ref{fig:textures}(b), a
geometry that is also highly interesting for Sm-$C$ elastomers.  In
each case the sample is effectively held by rigid clamps  -- tape
overlapping the sheet sample, and by glass plates in the slab case.
We expect changes in the polarization, $\vec{P}$ to be along $\y$
and hence suitable flexible electrodes would be needed for detection
of charges on the $xz$ faces for the sheet sample. In the slab case,
rigid $xy$ electrodes under the $xy$ plates are suggested for
detecting $xz$ surface charges by capturing the lines of $\vec{E}$
emerging from the sample surfaces with $\vec{y}$ as normal.

The clamps in each case eliminate in their vicinity $\lambda_{xx}$
and also  $\lambda_{yy}$ in the slab case.  For sheets this
constraint is over a small fraction of the sample.  The variation of
$\lambda_{xx}$, say, in the $z$ direction away from the boundaries
to more favorable bulk values is slow, thus generating (by
compatibility) minor $\lambda_{xz}$ additions to the deformation.
Under these circumstances one could expect elastically the sheet
deformations to be of the soft form eqn~(\ref{eq:soft}) where the
additional freedom gives a much reduced energy cost. By contrast,
the slab geometry imposes a constant $\lambda_{xx} = 1$ throughout
because the sample is thin in the $z$ dimension. Slabs then could at
best have the distortion eqn~(\ref{eqn:softish}) where the diagonal
elements are all 1. Analogously in slabs, rigid $xy$ plates would
suppress in their close vicinity any $\lambda_{yx}$ shear. It is
possible that in the bulk of slabs $\lambda_{yx}$ is suppressed,
though $z$-variation of $\lambda_{yx}$ generates through
compatibility $\lambda_{yz}$ which we are already considering and it
is possible that even for slabs, in the bulk one must consider
$\lambda_{yx} \ne 0$.

In both sheets and slabs, we have seen how imposing $\lambda_{xz}$ induces macroscopic translations in the $y$
direction resulting from the accompanying $\lambda_{yx}$ and $\lambda_{yz}$ shears.  Such shears drastically reduce
the energy. However, when the rigid bounding plates are constrained to move only in the $x$ direction,
there is a  conflict with the sympathetic shears that can be
 overcome by the development of
microstructure or textures of bands of $yz$-shears of alternating
sign, corresponding to rotations of alternating sign, $\pm \phi$,
but with the same $\lambda_{xz}$. At the macroscopic level the
$\lambda_{ij}$ components that are odd in $\phi$ cancel with each
other, and those even in $\phi$ add. Thus a body, with $\pm\phi$
rotations and thus also $\pm$ shears in equal proportions, as a
whole has a deformation gradient of the form:
\bea
\ten{\lambda}\s{mac} = \left(
\begin{array}{ccc}
\lambda_{xx}&0&\lambda_{xz}\\
0&\lambda_{yy}&0\\
0&0&1
\end{array}
\right) \label{eqn:quasi}
\eea
even though locally $\lambda_{yx}(\pm \phi) \ne 0$ and $\lambda_{yz}(\pm \phi) \ne 0$. (The $xx$ deformation
gradient might vanish, $\lambda_{xx} = 1$, as discussed above for slabs.)  The alternation ensures no macroscopic,
constraint-violating, $\y$-displacements develop across the sample. Such textures will be required for all three
deformations we have explored, soft or non-soft. The development of microstructure is analogous to that in the soft
deformations of nematic elastomers where the underlying soft deformations are observed, but where necessary in
textures to allow the underlying soft response to occur \cite{Zubarev_stripes:99}. Texture is not the cause of
softness, but a symptom that arises from softness in the presence of constraints.

The reader can see a concrete example of how to construct
$\ten{\lambda}\s{mac}$ by putting, say, the $\phi = \pi/2$ body of
fig.~\ref{fig:soft_modes} on top of that with $\phi = -\pi/2$.  (At
this special $\phi$, this picture is close to that in
Fig.~\ref{fig:softishtexture} below.) Looking at the position of the
overall top $xy$-surface with respect to that of the bottom, it is
clear that there has been a $xz$-shear, but that overall there is no
$y$-relative displacement and thus overall no $\lambda_{yz}$. What
is important is that the bodies are placed on top of each other,
rather than in any other configuration, a question of the
compatibility of textures to which we turn below. With the above
$\ten{\lambda}\s{mac}$ the energy will have been reduced (to zero in
the ideal soft case) and yet overall the body has conformed to
boundary conditions that are inconsistent with the underlying
deformations (\ref{eqn:simple_hard}), (\ref{eqn:softish}) and
(\ref{eq:soft}).  The development of textures to achieve soft
response in the face of incompatible boundary conditions
quasi-convexifies the free energy, a process well-understood in
Martensite \cite{Bhattacharyabook:03}, and in nematic elastomers
\cite{Conti:02,warnerbook:07}. It is important in the response of
the constrained samples that we are studying here.

We now derive the full form of the textures that arise for the two
geometries that do not rotate the smectic layer normals.  We limit
ourselves to such textures since then the $xy$ sample faces can then
retain their orientation as deformation proceeds. Recall that the
textures we have described, with unrotating smectic layers,  are
only part of one of a possible two families of textures in Sm-$C$
elastomers \cite{Biggins_Bhattacharya:09}.

\subsection{Forms of textures}
Neighboring laminates in a texture with deformations
$\ten{\lambda}_{\pm \phi}$ that are in contact through a common
surface must suffer deformations that are rank-one connected
\cite{Bhattacharyabook:03}, otherwise there is a geometric
inconsistency between the specification of translations of their
interfacial elements by $\ten{\lambda}_{+\phi}$ and
$\ten{\lambda}_{-\phi}$, that is:
\be \ten{\lambda}_{+ \phi}  - \ten{\lambda}_{- \phi} = \vec{a}
\otimes \vec{s}_{\phi}= 2 \left(\!
\begin{array}{ccc}
0&0&0\\
\lambda_{yx}&0&\lambda_{yz}\\
0&0&0
\end{array}
\!\right)   \label{eqn:rank1} \ee
since $\lambda_{yx}(\phi)$ and $\lambda_{yz}(\phi)$ are both odd about
$\phi = 0$ whereas $\lambda_{xz}(\phi)$, $\lambda_{xx}(\phi)$ and
$\lambda_{yy}(\phi)$ are even. Here $\vec{a}$ is a vector in a
laminate's surface in the target state and $\vec{s}_{\phi}$ is the surface
normal of a laminate back in the reference state, that is the normal
to the boundary of a region that will, after rotations of $\pm \phi$, transform into a laminate. It is
straightforward to see that (uniquely)
\be \vec{a} = (0,1,0)\; {\rm and} \;\;
\vec{s}_{\phi} \propto (\lambda_{yx},0,\lambda_{yz}).\label{eq:normalvector} \ee
  As the textures
evolve with $\phi$, their laminates have normals in the $zx$ plane and, if they
rotate at all with changing $\phi$, it is about the $y$ axis. The
texture normal in the target space, $\vec{s}'_{\phi}$, is given by the usual
transformation for normals of planes embedded in an elastic solid,
$\vec{s}'_{\phi} = \ten{\lambda}_{\phi}^{-T} \cdot \vec{s}_{\phi}$.

\subsubsection{Textures without elongations and in-plane shears} 
\begin{figure}[!htb]
\begin{center}
\includegraphics[width=0.4\textwidth]{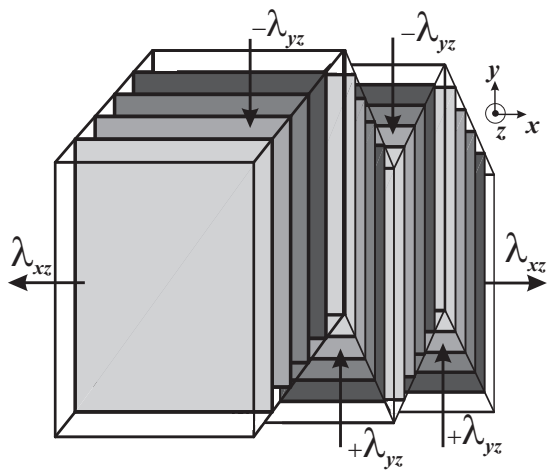}
\end{center}
\caption{
  A slice through a solid deforming with a textured version of
  (\ref{eqn:simple_hard}) for $\ten{\lambda}$.  Alternating $\pm
  \lambda_{yz}$ shears lead to no overall macroscopic $\lambda_{yz}$
  but to the desired imposed $\lambda_{xz}$.}
\label{fig:softishtexture}
\end{figure}

For the simple deformation eqn~(\ref{eqn:simple_hard})
with $\lambda_{yx} = 0$, the normal is $\vec{s}_{\phi} = \vec{s}'_{\phi} \propto
(0,0,\lambda_{yz})$. Thus the textures and the smectic layers share
the unchanging normal $\vec{s'} = \k_0 = \z$,
Fig.~\ref{fig:softishtexture}.
Successive $xy$ slices of the solid suffer alternating
$\lambda_{yz}$ transverse shears.  This pattern is a possibility for
slab geometries, but less likely for sheets where a soft texture is
a possibility we examine below.  There is no question of
polarization charges residing on the internal surfaces of this
texture since $\vec{P}$ rotates in the plane of the laminates and
never passes through  the internal interfaces.  This texture is a
special case of the generally charged family.

\subsubsection{Textures with in-plane shears}
\begin{figure}[!htb]
\begin{center}
\includegraphics[width=0.45\textwidth]{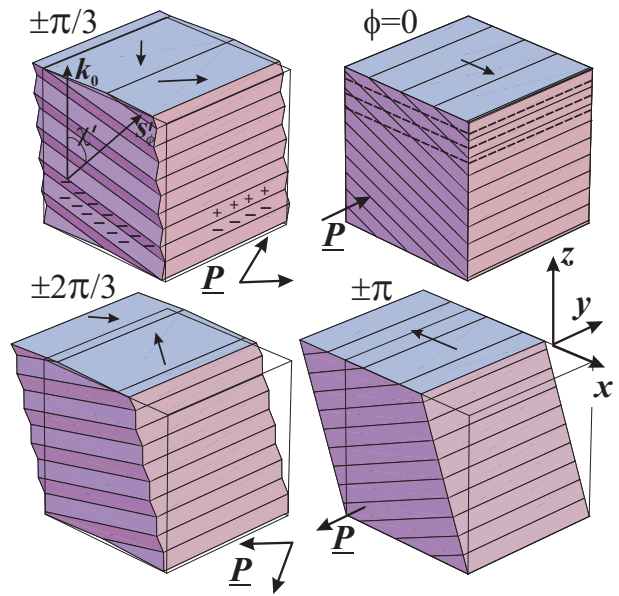}
\end{center}
\caption{An initially cubic sample  deforming softly with laminates of $yx$ and $yz$ shears of
  alternating sign and with $xz$ shear advancing from $0$ to
  $-2\Lambda$ in steps of $\pm \pi/3$ in director rotation $\phi$  about the $z$ axis
  (marked on figures; see also the $\vec{c}$ relevant to each laminate as it emerges at the top face of the sample). The smectic layers (a few dotted on the $\phi = 0$ snapshot) are unrotating as $\phi$ changes and retain their normal $\k_0 = \vec{z}$.  The alternating shear
  deformations lead to no corresponding macroscopic shears whereas the
  imposed $\lambda_{xz}$ does. The laminate normal $\vec{s}'_{\phi}$ in the deformed (target) state is shown
  for the example of $\phi = \pi/3$.  It starts and finishes parallel
  to $\n$ but in general makes an angle $\chi$ with the layer normal
  $\k_0$ given in the text. The deformations away from the initial
  shape (light reference frame given in $\phi \ne 0$ pictures) reveal
  the slight contraction $\lambda_{xx}$ along $x$ and a compensatory
  lengthening along $y$, see e.g. $\phi = \pi/3$, for $\phi \ne 0, \pm
  \pi$. The $\pm \pi/2$ case is like Fig~\ref{fig:softishtexture}, that is with $\chi = 0$, but
  with $\lambda_{xx},\lambda_{yy} \ne 1$. } \label{fig:softsequence}
\end{figure}

For both the other choices of deformation, the normal $\vec{s}'_{\phi}$ of
the transformed laminates has a universal form from letting the
appropriate $\ten{\lambda}^{-T}$ act on planes with normal $\vec{s}_{\phi}$ (eqn~(\ref{eq:normalvector})), see fig~\ref{fig:softsequence}.
The laminate orientation  is
independent of softness or not, that is it holds for both classes
(B) and (C) of section~\ref{sect:model}:
\bea \vec{s}'_{\phi}  &\propto &
  \left( \lambda_{yy} \lambda_{yx}, \;0, \;\lambda_{yz} -\lambda_{yy} \lambda_{yx} \lambda_{xz}\right)
 \label{eq:target_gen} \\
 & \rightarrow & ( \sin\theta \cos\phi, \;0, \; \cos\theta). \label{eq:target_universal}
\eea
The current (target space) laminate normal $\vec{s}'_{\phi}$ has an angle,
$\chi'$, to the $z$ axis:
\be
\chi'_{\phi} = \tan^{-1}
\left[\cos\phi\tan\theta\right].\label{eq:anglechi}
\ee
Contrast this case with the more trivial case above where the
laminate normal was along $z$, that is along the layer normal.  The
need for more subtle laminates can be seen in
fig.~\ref{fig:soft_modes} where, except for $\phi = 0, \pm \pi/2,
\pm \pi$, the top ($xy$) surface is sheared away from its initial
square shape with shears $\pm \lambda_{yx} \ne 0$ of alternating
sign. Now stacking $\pm \phi$ bodies on top of each other to
eliminate a macroscopic $\lambda_{yz}$ shear is impossible since the
faces do not match, unless the laminates are tilted with respect to
the layers.

 For different rotation angles $\phi$ of
the $\c$-director around $\k_0$, the normal $\vec{s}_{\phi}$ or $\vec{s}'_{\phi}$ to the laminates
takes differing rotation angles about the $\y$ axis, Fig.~\ref{fig:softsequence}.  For $\phi = \pi/2$, one has $\chi
\chi' = 0$,  that is $\vec{s}_{\phi}$ and $\vec{s}'_{\phi}$ are along $\k_0$ and the
laminates are aligned with the smectic planes.  Again
fig.~\ref{fig:soft_modes} makes this clear -- at this $\phi$ there
is no $yx$ shear and no impediment to simply stacking the $\pm$
regions as simple laminates.
To explain why rank-one connectedness requires laminates rotated with respect to the smectic layers, Fig.~\ref{fig:softsequence}, we transform $\ten{\lambda}$ from the $(x,y,z)$ frame to one based upon $\vec{s}$, that is $(\perp, y, s)$ by rotating both the target and reference states, thus $\ten{\lambda} \rightarrow \ten{\lambda}' = \ten{W}\cdot\ten{\lambda}\cdot\ten{U}^T$ by $\chi$ about $y$.  The target and reference state rotations are $\ten{W}$ and $\ten{U}$ respectively.  The rotations have the effect of mixing the $\lambda_{yx}$ and $\lambda_{yz}$ shears to give a $\lambda_{y\perp} = \cos\chi \lambda_{yx} - \sin\chi\lambda_{yz}$ which vanishes by definition of $\vec{s}$, eqn~(\ref{eq:normalvector}).  The laminate planes at this orientation no longer suffer the in-plane shears that would make them impossible to match across their interfaces.

Our analysis of soft modes and non-soft modes with $yx$-relaxation,
has been based upon the free rotation of the normal of the system of
laminates about the $y$ axis.  It is possible that the motion of the
laminate surfaces, through the assembly of smectic layers that is
stationary, is in fact pinned. A rubber is liquid-like at the
molecular level, accounting for the extreme extensibility of rubbers
and, for instance, their motionally-narrowed NMR lines.  It is known
that laminates evolve and rotate easily in nematic elastomers
responding at low energy cost \cite{Zubarev_stripes:99}. Experiments
on polydomains \cite{Sanchez:08} show that is is likely that
textures in Smectic C elastomers can evolve under imposed strains
and that low-energy deformations are thus possible as we propose.
Experiment is vital to determine whether pinning is active or not.

Another uncertainty is the energy cost of laminate walls.  The
directors on opposite sides of a laminate surface are rotated to
$\pm \phi$.  At $\phi = \pi/2$ the layers and laminates share a
common normal and laminate surfaces are a $\pi$ twist-bend  wall. At
other values of $\phi$, the laminates and layers are oblique to each
other and the wall is more complicated.  In any event there is a
cost to walls that demands calculation when experiment in the two
geometries has been carried out. This energy cost and any competing
one associated with the texture will determine, as in nematic
elastomers, the ultimate length scale of the structure.  A simple
qualitative argument yields the usual scaling form of the texture
periodicity: Consider a simple case where the texture is $xy$ slabs
of thickness  ${\cal L}$ and with surface energy $\gamma$ per unit
area. Let $w_i$ be the extent of the sample in the $i^{\rm th}$
direction. Then there are $w_z/ {\cal L}$ slabs each of energy
$\gamma w_x w_y$. Looking ahead to our analysis of polarization,
there are  strips of surface charge alternating with period ${\cal
L}$ on the $yz$ ends of the sample, see
figures~\ref{fig:softsequence}.  These charges give fields extending
outside the sample an $x$-distance $\sim {\cal L}$ (by Poisson's law
they decay exponentially in the $x$ direction with decay length
${\cal L}$). The fields scale with $P \cos\phi$ and thus the field
energy outside the sample is of the form $\gamma_p {\cal L} w_y w_z$
where $\gamma_p \sim \epsilon_0 P^2$. The overall energy has
competing ${\cal L}$ terms. Minimization over ${\cal L}$ gives
${\cal L} \sim \sqrt{w_x \gamma/ \gamma_p}$.  More complicated
calculations are needed for greater precision.

\subsubsection{Disproportionation}
\begin{figure}[!htb]
\begin{center}
\includegraphics[width=0.35\textwidth]{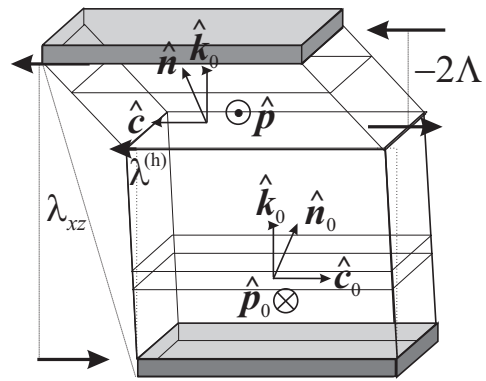}
\end{center}
\caption{
  A sample deforming by disproportionation.  The macroscopic
  $\lambda_{xz}$ shear displacements are indicated by heavy arrows and
  consists of a fraction (lower) of weakly sheared ($\lambda^(h)$) sample and the complementary (upper)
  fraction distorted by $-2\Lambda$.  The macroscopic shear is $\lambda_{xz}$.} \label{fig:disproportionate}
\end{figure}

One perhaps important possible form of texture
remains, a simple disproportionation that might occur in both slab
and sheet geometries.  Analogous disproportionations have been proposed in nematic elastomers \cite{Conti:02b}, though in practice it seems textures are selected.  The top portion of the sample is $xz$-sheared
to its maximum soft extent, $-2\Lambda$, with rotation $\phi = \pi$ and the bottom portion is weakly $xz$ sheared by $\lambda^{(h)}$ with unrotated director, see
Fig.~\ref{fig:disproportionate}.
The bottom must be sheared so that it has a shear stress $\sigma_{xz}$ to match that in the upper part.  Any discontinuity of this stress on traversing the interface between the two parts would lead to a net body force.

If the material were ideally soft then the shear stress in
distorting the upper portion would be zero and the lower portion
would thus remain unsheared. The two states would be separated by a
$\pi$ twist-bend wall. It is possible that a sample might deform
this way if rotations of textures are pinned and if the energy cost
of laminate interfaces discussed above is too high.  Only one
interface is required for this mode of deformation, and that
interface's normal does not rotate with increasing over all
$\lambda_{xz}$. Simple geometrical considerations determine the
volume fraction $\Phi$ of sample that is transformed into the $\phi
= \pi$, $\lambda_{xz} = -2\Lambda$ state.  For this soft case it is
$\Phi = \lambda_{xz}/(-2\Lambda)$. Thus the interface migrates down
the sample of figure~\ref{fig:disproportionate} as $\lambda_{xz}$
increases.  A particularly simple ferro-electric response then
arises, as we see in the next section, and no internal surfaces are
charged and thereby add to the energetic cost.

In practice there are small, so-called ``semi-soft'' elastic energies
to be paid in non-ideal elastomers where softness would otherwise
hold.  Our current smectic elastomer model is that of nematic
elastomer distortion energy under smectic layer spacing constraints.
In nematic elastomers non-ideality is known to introduce a threshold
before essentially soft deformation associated with director
rotation on sets.  There is a plateau where the stress increases
only slowly until the director fully rotates, during which the
sympathetic director rotation and the elongations, contractions and
shears are identical in form to the ideal case \cite{warnerbook:07}.
For  non-ideal  Sm-$C$ elastomers disproportionating, the commons
stress in the two portions is determined by the upper portion is
that characteristic of the end of the semi-soft deformation plateau,
that is the maximum value of low stresses before hard deformation
starts.  To develop a matching stress, the bottom portion would then
have to $xz$ shear as well, without any sympathetic deformations
(since they would not be rank-one connected to the state of the
upper portion).  This would be a hard deformation and thus of small
amplitude $\lambda^{(h)}$, as sketched in the
figure~\ref{fig:disproportionate}.  Now the volume fraction $\Phi$
that is transformed into the $\phi = \pi$ state  is $\Phi =
(\lambda_{xz}-\lambda^{(h)})/(-2\Lambda+\lambda^{(h)})$ and  the
interface would migrate down the sample as $\lambda_{xz} \rightarrow
-2\Lambda$.

However, because of non-ideality in practice such simple
disproportionating response would not seem probable -- the
alternative is to form textures in the way we have outlined above.
One would initially find lower stresses as the shear develops,
characteristic of the start of the semi-soft plateau associated with
the $\pm \phi \sim 0$ laminates.  In fact  at any given
$\lambda_{xz} > - 2\Lambda$ uniform through the sample, the stresses
would be less than in the disproportionating scheme.  Textures would
therefore develop at lower energy cost than the higher stress
alternative of developing the same $\lambda_{xz}$ through
disproportionation. The favoring of textures over disproportionation
would be lost as samples approach extreme softness and the cost of
internal boundaries gets relatively high.

\subsection{An alternative route to observing microstructures}

We have proposed experimentally observing textures in Sm-$C$
elastomers from thin samples in the sheet geometry, and polarization
microscopy. Classical buckling instabilities are eventually observed
in shear experiments of this sort and may obscure the textures
\cite{Kramer:08,Stenull:08}. Alternatively, a tensile geometry
provides a simpler experiment to reveal soft deformations in Sm-$C$
elastomers, and their associated microstructures.
\begin{figure}[!htb]
\includegraphics[width = 0.49\textwidth]{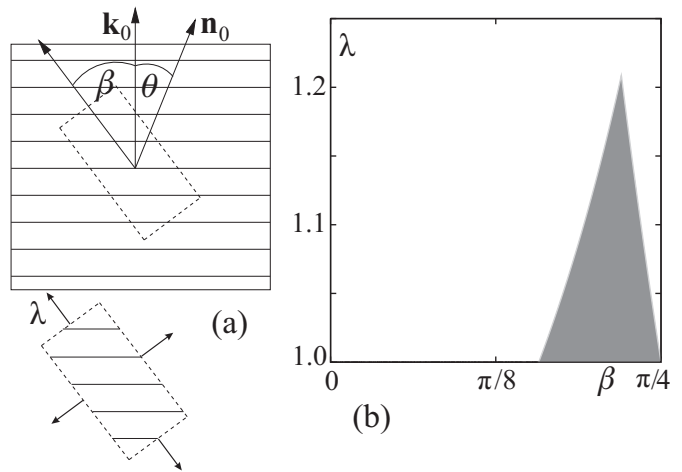}
\caption{(a) A sample with the layer normal at an angle $\beta$ to the long
  direction is cut out of a monodomain Sm-$C$. (b) The shaded region is that of elongations $\lambda$
 for tension angles $\beta$ for which the sample remains soft  (for $\theta = 22.5^\circ$ and $r=2$).}
\label{fig:tensileexp}
\end{figure}
A sample of elastomer with an arbitrary angle $\beta$ between the
proposed elongation axis and the layer normal can be prepared by
cutting from a larger sample as illustrated in
Fig.~\ref{fig:tensileexp}(a). When this sample is stretched then it
will deform softly provided it can form appropriate microstructures \cite{adams:06a}.
The extent of the soft deformation is illustrated in
Fig.~\ref{fig:tensileexp}(b).  For small $\beta$ there are no
microstructures that permit a soft deformation, but for  an interval of $\beta$ from above $\pi/8$ until
close to $\pi/4$, the sample can deform softly.

\section{Ferro-electric response}
When a polarization $\vec{P}$ changes on passing through a surface, by Gauss's
theorem a surface charge density $\sigma = -\Delta P_{\perp}$ develops
that is equal to the change of the normal component of $\vec{P}$ from
one side to the other. Thus at the $xz$ external surfaces of both sample
geometries, $\sigma^{(xz)} = P \cos\phi \equiv P_y$;  see for instance the explicit decoration of $xz$ surfaces with charges in the $\pm \pi/3$ snapshots in figure~\ref{fig:softsequence}.  We now
relate $\sigma^{(xz)}$ to the imposed deformation $\lambda_{xz}$ in the three cases of deformation.

\noindent (i) The non-soft deformation with a threshold has singular director
rotation against imposed $\lambda_{xz}$ because it too is essentially
of the $\cos^{-1}$ form that subsequent deformation modes are also
shown to be.  Between $\lambda_1$ and $\lambda_2$, the variation is
almost indistinguishable from $\cos\phi = (2\lambda_{xz}
-\lambda_1-\lambda_2)/(\lambda_1-\lambda_2)$ where $\lambda_1$ and
$\lambda_2$ are the thresholds to director rotation that arise in this
constrained case.  Thus the surface charge evolves linearly with
$\lambda_{xz}$:
 \be \sigma^{(xz)} = -P (2\lambda_{xz}
-\lambda_1-\lambda_2)/(\lambda_1-\lambda_2). \label{eq:hardpol}
\ee

\noindent (ii) For the non-soft deformation without a threshold,
eqn~(\ref{eqn:softish}), one sees directly from
eqn~(\ref{eq:softish-xz}) that the $xz$-surface charge density is:
\be \sigma^{(xz)}(\lambda_{xz})  = -P
(1+\lambda_{xz}/\Lambda). \label{eq:softishcharge}
\ee
Recall that the imposed shear $\lambda_{xz}$ varies from $0$ to
$-2\Lambda$ and thus this surface charge reverses linearly with the
applied shear deformation. It would not change any further, were the
shear to be increased beyond $-2\Lambda$.

\noindent (iii) When deformation is soft, the connection between $\cos\phi$ and
$\lambda_{xz}$ is slightly more complicated, despite being linear at
the start and end of the rotation.  Eqn~(\ref{eq:softphilambdaxz}) for $\cos\phi \equiv \sigma/P$
gives the explicit shear-dependence of the $xz$ surface charge density:
\be \sigma^{(xz)}\s{soft}  = -P
\frac{r}{\rho}\hspace{-1mm}\left( \hspace{-1mm}\sqrt{\left(\frac{\rho}{r}\right)^2\! +\!
\left(\!1-\frac{\rho}{r}\!\right)\!\left(\frac{\lambda_{xz}}{\Lambda}\right)^2}\! +\!
\frac{\lambda_{xz}}{\Lambda} \hspace{-1mm} \right)\hspace{-1mm}.\label{eq:softcharge} \ee

\subsection{Charging of laminate surfaces; overall electrostatic energy and mechanical stability}
\label{sect:internal-charge}
For the non-trivial textures with the laminate normals not along the
layer normals, the normal component of the polarization can reverse at the internal surfaces
separating laminates in the texture. One must then consider the
energies of the resulting internal surface charge distributions in
both the cases of sheets and slabs.  The laminate  normals $\vec{s}'_{\phi}$
rotate about $y$ in the $xz$ plane.  The $y$-component of $\vec{P}$,
giving rise to the $xz$-external surface charges discussed above and
pictured in figures~\ref{fig:cartoon_C}, \ref{fig:textures} and the $\pm \pi/3$ part of \ref{fig:softsequence}, does not
contribute to the internal sheets of charge.  The internal charges arise rather from
the reversal of the $P_x$ component at the laminate surfaces, that is
$\sigma\s{lam} \propto P \x \cdot \vec{s}'_{\phi}$; see figure~\ref{fig:softsequence}. The internal surface
charge density is therefore
 \be \frac{\sigma\sp{lam}}{P}  = \pm \sin\phi \sin\chi = \pm \half \frac{\tan\theta \sin
2\phi}{(1 +\tan^2\theta \cos^2\phi)^{1/2}} .\label{eq:internal-charge}\ee
One must ask then whether these charges have a significant energetic
effect?

In the sheet geometry relatively little of the otherwise uniform
field generated by the $xz$-sheets of external surface charge leaks
out of the sample.  The $y$-internal electric displacement in the
sample is $D\sp{int}_{(y)} = \sigma^{(xz)} = -P\cos\phi $. If the
textures are fine, the electric displacement between internal
surfaces of alternating charge is also not lost, and is
$D\sp{lam}_{\vec{s}'_{\phi}}=P\sin\phi \sin\chi $, from
eqn~(\ref{eq:internal-charge}).  This $D$ is in the $\pm \vec{s}'_{\phi}$
directions.

  Finally the component of $P_x$ that  intersects the external $yz$-surfaces generates
$\sigma^{(yz)}$ external surface charge densities, see the charges shown on the $yz$ surfaces of the $\pm \p/3$ snapshot of
figure~\ref{fig:softsequence}. These are $\sigma^{(yz)} = -P\sin\phi
\cos\chi $ and change sign between the $yz$ ends of neighboring laminates as
$\phi \rightarrow -\phi$.
The charges generate small effective internal fields because the $yz$
surfaces are small, are widely separated in the sheet geometry and, in
any event, are oppositely charged in stripes, so we can ignore their
energetic effect.

At the exceptional points $\phi = \pm \pi/2$, the
internal fields $E\s{int}^y = 0$ and $E\s{lam} =0$, the latter because
the laminate normals are along the layer normals $z$.  All the surface
charges are confined to the $yz$ external surfaces which we have just
argued generate an ignorable internal field.

The overall
electrostatic energy density, $f\s{el}$, is thus essentially only from internal fields generated by charges appearing on the internal surfaces of textures.  These fields depend
in part on the anisotropic dielectric tensor of the elastomer which we
represent for simplicity, see appendix~\ref{app:electro}, by a single
averaged value $\epsilon$:
\bea  f\s{el} &=& \half \frac{P^2}{\epsilon \epsilon_0} \left\{ \cos^2\phi + \sin^2\phi \sin^2\chi \right\} \nonumber \\
  &=& \half \frac{P^2}{\epsilon \epsilon_0}  \cos^2\phi/\left(  \cos^2\theta + \sin^2\theta
  \cos^2\phi\right)\label{eq:electro_energy}. \eea
There is a weak $\chi$-dependence in $1/\epsilon$ that we estimate in
Appendix~\ref{app:electro} and argue that we can neglect in the
estimates below.

At first sight this energy is perhaps alarming.  $f\s{el}$ is maximal,
$f\s{el} = P^2/(2\epsilon \epsilon_0)$ at $\phi = 0, \pm \pi$ and is
minimal, $f\s{el} = 0$, at $\phi = \pm \pi/2$.  For soft deformations,
where there is no elastic cost, the director should spontaneously
rotate to $\phi = \pm \pi/2$ and the sample should spontaneously shear
to $-\Lambda$. It is likely however that the $xz$ external surfaces,
that are initially charged, attract counter-ions from their
surroundings and are neutralised.  There is accordingly no
electrostatic cost at $\phi = 0$ (and finally at $\phi = \pm \pi$)
since no internal field $E\s{int}^y$ is generated.  The only internal
fields are those between successive (internal) laminate surfaces
(unless these too are neutralised by internal conduction processes).
The electrostatic energy density is then simply:
\bea  f'\s{el} &=& \frac{P^2}{\epsilon
\epsilon_0} \frac{\sin^2 2\phi   \sin^2\theta}{ \cos^2\theta + \sin^2\theta
  \cos^2\phi} \label{eq:int_electro_energy} . \eea
There is thus no electrostatic cost at $\phi = 0, \pm\pi/2, \pm \pi$.
All these points are local minima of the free energy and cost in
deviating from these points means that spontaneous rotation is
avoided. It does suggest however that there is a complex energy
associated with the path between $\lambda_{xz} = 0$ and $-2\Lambda$
for softly deforming elastomers.  Should the deformation be of the
hard type with $\lambda_{yx} \ne 0$ which demands complex textures as
in the soft case, then the elastic energy has to be added to the
electrostatic energy $f\s{el}'$, the resultant of which depends on the
relative scale of rubber elastic and ferro-electric energies.

Values of $P$ for smectic C* systems are in the range
$10^{-5}-10^{-4}{\rm C/m}^2$ \cite{warnerbook:07}, with saturation
values of $P$ for some polymers used for Sm-$C$* networks being
reported as large as $2\times 10^{-3}{\rm C/m^2}$ \cite{Koehler:05}.
One must compare the resulting energy density, $\sim \half
P^2/\epsilon_0$, with elastic energy densities in the problem which
are $\half \mu 10^{-2}$, taking the strain energy at the maximal
(barrier) value, $10^{-2}$ in units of $\mu/2$, see
Fig.~\ref{fig:hard_deformation}(a). For $\mu \sim 10^{5}{\rm J/m^3}$
this puts the ratio of the electro-static to elastic energies in the
range $10^{-2}$ to $1$ -- under some circumstances electrostatics
may be important, making the need for experimental results still
greater to discern the different mechanisms that will be selected by
these competitive energies.  It is possible that one might have
complicated free energies of deformation that result from the sum of
the elastic and electrostatic influences.

Slabs would appear to have the choice of hard deformations with or
with out the additional $yx$ shear relaxation.  Without $yx$ shears,
the laminate normals are along the layer normals and hence always
perpendicular to the polarization that rotates about the $z$ axis as
$\phi$ evolves.  Accordingly, no internal sheets of charge develop for
intermediate imposed shears.  With $yx$ shears the laminate surfaces
have normals as in the soft case and the same internal surface charges
are generated.  If the texture is fine, the distance between
alternating internal sheets of charge is small and little of the
associated field leaks out.  Then there would be an energetic cost as
for laminates discussed above.

The charges on the $xz$ and $yz$ external surfaces give rise to
fields that largely leak out of the sample and hence again we ignore
their contribution to the energy.  The leaking field lines from the
$xz$ surfaces can be captured by overhanging electrodes as in
Fig~\ref{fig:textures}(b) and hence the surface charge, and $P$, can
be measured.

As mentioned above,  we have described only one of a possible two
families of textures in Sm-$C$ elastomers.  The other family of
textures can be shown \cite{Biggins_Bhattacharya:09} not to have
charges accumulating on their internal surfaces.  They are not
necessarily applicable to the constraints of unrotating smectic
layers that we have considered.

\subsection{Electrically driven actuation?}
\label{sect:electro-activation} The inverse response when applying a
potential to elastomers in the two geometries in order to generate a
shear strain also needs quantifying.  Elastic barriers to
deformation still exist, but now one is coupling external potentials
to the charge layers that exist on the free surfaces, that is
achieving an energy change per unit $xz$-area of sample of $\half
\sigma V$ on achieving $\sigma = 0$ as a result of applying a
voltage $V$ across the sample and mechanically switching as far as
the maximum in the elastic energy barrier (where $\phi = \pi/2$).
For the favorable case of the sheet, taking $V=10^3$V and thickness
$w = 1$mm, the relative energy densities (electrical to elastic) are
$V P/(w \mu 10^{-2}) \sim 10^{-2}$ -- 1 considering the range of
values of $P$ given above. The reason is that internal electrical
fields that set the scale of energies above are $E\s{int}\sim
10^{7}{\rm V/m}$, much larger than those typically able to be
applied here. One concludes that electro-mechanical actuation at
very large shear strains ($\sim 1$) will require the sheet geometry with large
fields.

\subsection{Experimental observations}
The internal fields, that we argue could play a significant role in
fixing the mechanical stability of these elastomers in their
textured state, are perhaps open to direct observation that might
help determine what rotations are accompanying shears.  These
intense fields are given, see Appendix~\ref{app:electro}, by
$\vec{E} =\frac{1}{\epsilon_0}\ten{\epsilon}^{-1} \cdot \vec{D}$
with $\vec{D} = P_{s'} \vec{s}'_{\phi}$ the electric displacement generated
by the internal surface charges and directed along the laminate
normals.  Thus $E_i =\frac{1}{\epsilon_0}\epsilon^{-1}_{is'}
D_{s'}$. The field is not purely along the laminate normal because
this direction is not a principal direction of the dielectric
tensor.  Guest species with, for instance, characteristic absorption
or fluorescence could be aligned with the field rather than with,
say, the director or smectic layer normal.  In particular a
distinction between the field and director directions emerges as
soon as rotation starts and would be an independent check of the
development of microstructure.  Absorption or florescence probes
suggested above are invariant under $\vec{s}'_{\phi} \rightarrow -\vec{s}'_{\phi}$
and thus would not find the alternating texture structure nugatory.
It is likely that the choice of geometry, slab or sheet, will yield
very different results.

\section{Conclusions}

The deformation path taken by a smectic-C elastomer in having the
direction of its spontaneous polarization mechanically or
electrically rotated and eventually reversed is calculated for slab
and sheet geometries.  We are concerned with the Goldstone mode of
director rotation about layer normals on a cone of fixed angle,
rather than mechanical and electrical induction of a change in cone
angle.  Experiments on the alignment of polydomain  smectic
elastomers \cite{Sanchez:08} by external stress demonstrate that
director rotation is easily possible and is an important aspect of
deformation, as it is also in nematic elastomers.  To conform to
constraints while deforming at low elastic energy, it is proposed
that textures develop.  The laminates of these textures are at
non-trivial angles to the layer system, and rotate with respect to
the layers at the strain develops.  Different possible textures
result according to geometry, the constraints at surfaces, the
elastic shear energy, the role of semi-softness, and the charging of
internal texture surfaces when they cut the polarization.  One
extreme case of geometric disproportionation is discussed and found
to be unlikely.  However, important determinants such as the
relative roles of polarization and elastic cost are still most open
and experiments on large deformations are urgently required.

\section*{Acknowledgements} We are grateful to Prof. K.
Bhattacharya for pointing out the necessity to examine the effect of
charges on internal surfaces that led to the analysis of
section~\ref{sect:internal-charge}.  John Biggins made important
suggestions about disproportionation and on the character of Sm-$C$
stripe-domains generally. Hiraoka {\it et al} gave us permission to use the
photographs of figure~\ref{fig:cartoon}.

\appendix
\section{Energy scale $\mu$ for deformations}\label{app:energy}
We model smectic elastomers as essentially nematic elastomers with
strong constraints of layering.  Thus $\mu$ that scales the underlying
rubber elastic energy would be the shear modulus for the rubber, were
it able to enter its isotropic state.  More usefully one can relate it
to certain shear or extensional moduli in the Sm-$C$ elastomeric state.
Elongation along the $\c_0$ direction, $\lambda_{xx} >1$, or shear
with displacement in the positive $x$ direction, $\lambda_{xz} >0$,
both do not rotate the $\c$-director, provided the layer normal $\k_0$
is fixed (for instance by clamps or rigid plates).  We are assuming
rigid Sm order -- for instance elongations are accompanied by
$\lambda_{yy} = 1/\lambda_{xx}$ since contraction along $z$ is
forbidden (it would alter the layer spacing) and these deformations
are two-dimensional as in the case of Sm-$A$ elastomers undergoing
similar strains \cite{Nishikawa:99} and which have been modelled in
these terms \cite{adams:05}. Without rotations and other shears, the
free energy cannot drastically reduce, as in the rest of this
investigation.  The free energies of such distortions are:
\bea f(\lambda_{xz} \ge 0) &=& \half \mu \left( 3 + \frac{1}{4r}
\left[ r+1 + (r-1) \cos2\theta \right]^2\lambda_{xz}^2 \right) \nonumber\\
f(\lambda_{xx} \ge 1) &=& \half \mu \left( \frac{(\lambda_{xx}-1)^2}{8r} \left[ r^2+1  - (r-1)^2
\cos4\theta \right] \right. \nonumber \\ && \left. +  \frac{1}{4\lambda_{xx}^2} \left[ 4 +
3\lambda_{xx}^2 +2\lambda_{xx}^3 +3\lambda_{xx}^4\right]\right) \nonumber
\eea
with corresponding small strain moduli:
\bea k_{xz}  &=& \frac{\mu}{4r}
\left[ r+1 + (r-1) \cos2\theta \right]^2 \nonumber\\
k_{xx}   &=& \frac{\mu}{8r} \left( 1+ 30 r +r^2 - (r-1)^2
\cos4\theta \right). \nonumber \eea

\begin{figure}[!htb]
\begin{center}
\includegraphics[width=0.45\textwidth]{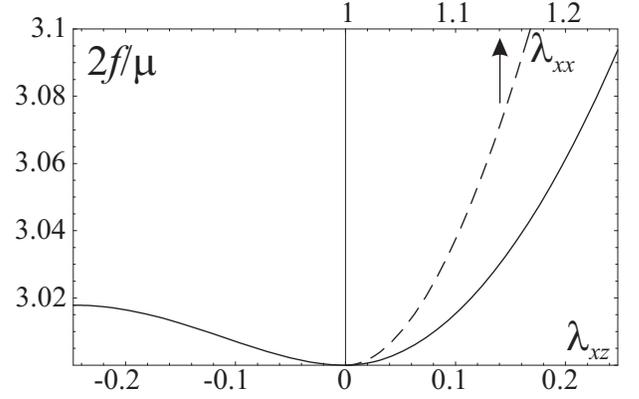}
\end{center}
\caption{
  Energy of deformation for both positive (hard) and negative (softer)
  shear $\lambda_{xz}$, and for extensions $\lambda_{xx} \ge 1$
  (dotted, upper scale) up to strains of $\pm \Lambda = \pm 0.246$, for the values of $r$ and $\theta$ adopted in the text.  Note how much
  lower the energy of distortion is if sympathetic relaxations and
  director rotation are permitted.} \label{fig:energy_scale}
\end{figure}

A sense of how much easier deformation is with relaxation can be seen
in Fig.~\ref{fig:energy_scale}.
The moduli, along with the spontaneous distortion (or limits to softer
deformation and completion of charge switching) and conoscopy (for
$\theta$) give experimental insight into $\mu$, $r-1$ and
$\cos\theta$.

\section{Non-linearity and apparent asymmetries in deformations}\label{app:asymm}

The deformation (\ref{eqn:softish}) leads to a $\lambda_{yz}$
apparently asymmetric about $\phi = \pi/2$, or equivalently
$\lambda_{xz} = - \Lambda$.  (A dotted vertical line in fig.~\ref{fig:softish_deformation} at $\lambda_{xz} = -\Lambda$ makes clear the lack of symmetry.)  Manifestly when $\c$ has been rotated to
$\y$, the $yz$ shear is geometrically maximal and should reverse to
zero as $\phi \rightarrow \pi$ just as it advanced while $\phi
\rightarrow \pi/2$.  We demonstrate that the failure of
(\ref{eq:softish-yz}) and Fig.~\ref{fig:softish_deformation} to
exhibit symmetry about $\lambda_{xz} = - \Lambda$ is only apparent.
It is a consequence of geometrical non-linearity (the need to compound
rather than add finite deformations).

Consider deformed states of the solid at $\phi = \pi/2 + \zeta$.  One
can compound deformations:
\be  \ten{\lambda}_{\phi} = \ten{\lambda}_{\zeta}' \cdot \ten{\lambda}_{\frac{\pi}{2}} \nonumber
\ee
where $\ten{\lambda}_{\phi}$ and $\ten{\lambda}_{\frac{\pi}{2}}$
(unprimed tensors) are with respect to the original reference state,
$\vec{X}^0$ say, while $\ten{\lambda}_{\zeta}'$ is a deformation
gradient with respect to the reference state $\vec{X}'$ resulting from
the deformation $\ten{\lambda}_{\frac{\pi}{2}}$, that is $\vec{X}' =
\ten{\lambda}_{\frac{\pi}{2}}\cdot \vec{X}^0$.  Using eqns~(\ref{eq:softish-xz})-(\ref{eq:softish-yx}) with $\phi = \pi/2$ for  $\ten{\lambda}_{\frac{\pi}{2}}$, and adopting the form
(\ref{eqn:softish}) again for
$\ten{\lambda}_{\zeta}'$, one obtains respectively:
\bea \left(
\begin{array}{ccc}
1&0&-\Lambda\\
0&1&\Lambda\\
0&0&1
\end{array}
\right) \;\;{\rm and} \;\; \left(
\begin{array}{ccc}
1&0&\lambda_{xz}'\\
\lambda_{yx}'&1&\lambda_{yz}'\\
0 &0&1
\end{array}
\right) .\nonumber    \eea
In $\ten{\lambda}_{\zeta}'$ one has $\lambda_{xz}'$ odd in $\zeta$ since it continues to decrease below $-\Lambda$. Likewise
$\lambda_{yx}'$ is also odd since it has returned to zero at $\zeta = 0$ and must now become negative for the second half of the range of $\phi$.  The shear  $\lambda_{yz}'$ is even since it reached its maximum of $\Lambda$ when the director points transversely and then must decline again.  The roles of $x$ and
$y$ in shears involving $z$ have thus been interchanged in this reference
state for $\ten{\lambda}'$ with $\c$ initially along $\y$, just as it
was initially along $x$ for the reference state of $\ten{\lambda}$.
Important for this argument is that rotations of $\pm \zeta$ are
physically equivalent.

Consider the difference $\ten{\Delta\lambda}$ in the
$\ten{\lambda}_{\frac{\pi}{2} + \zeta}$ and
$\ten{\lambda}_{\frac{\pi}{2} -\zeta}$ tensors --- it will expose any
asymmetry in $ \ten{\lambda}$ about the $\lambda_{xz} = - \Lambda$
point.
\bea \ten{\Delta\lambda} &=& (\ten{\lambda}_{\zeta}'-\ten{\lambda}_{-\zeta}'  )\cdot
 \ten{\lambda}_{\frac{\pi}{2}}\nonumber \\
  &=& 2 \left(
\begin{array}{ccc}
0&0&\lambda_{xz}'(\zeta)\\
\lambda_{yx}'(\zeta)&0&0\\
0 &0&0
\end{array}
\right) \cdot \left(
\begin{array}{ccc}
1&0&-\Lambda\\
0&1&\Lambda\\
0&0&1
\end{array}
\right)\nonumber \\
  \Delta\lambda_{yz} &=& - 2\lambda_{yx}'(\zeta)\Lambda \ne 0 \nonumber
\eea
The extra, $yx$, relaxation permitted in (\ref{eqn:softish}) has
created an asymmetry in $\lambda_{yz}$ about $\pi/2$ that is not a
reflection of the true symmetry about $\phi = \pi/2$ that exists in
$yz$ shear. We used in the $\lambda_{yz}'$ this very (odd) symmetry with
respect to the $\phi = \pi/2$ state. [Likewise in (\ref{eq:soft}) for soft deformations,
allowing further additional elements in $\ten{\lambda}$ introduces
further geometrical non-linearity.]  The energy in the softer case however remains
symmetric about $\lambda_{xz} = -\Lambda$, as expected; see the dotted curve in fig.~\ref{fig:hard_deformation}(a).

\section{Electrostatic energy from internal fields.}\label{app:electro}
The laminate surface charge densities are $\sigma\sp{lam} = -\Delta
P_{s'}$ which generate the displacement fields $D_{s'} =
\sigma\sp{lam}$.  The electric field energy density is $\half
\vec{D} \cdot \vec{E}$ with $\vec{E}
=\frac{1}{\epsilon_0}\ten{\epsilon}^{-1} \cdot \vec{D}$.  From this
expression and given that $\vec{D}$ is in the $\vec{s}'_{\phi}$ direction,
it is clear for the energy that we need the element
$\left(\ten{\epsilon}^{-1}\right)_{s's'}$ that in
eqn~(\ref{eq:electro_energy}) and thereafter we have denoted by
$1/\epsilon$:
\bea f\s{el} = \frac{1}{2\epsilon_0}D_{s'}\epsilon_{s's'}^{-1}
D_{s'} \label{eq:appenergy}
\eea
(no summation over $s'$).

One can make various  estimates of the relevant element. One
assumption would be to take $\ten{\epsilon}$ to be uniaxial about
the director $\n$.  We have already made a similar assumption with
the $\ten{\ell}$ tensor.  However the Sm-$C$* phase is anything but
uniaxial in its electrical properties -- it has ferro-electric
ordering along the in-plane direction perpendicular to $\vec{c}$.
However, one could take the view that the ferro-electric ordering is
sterically-driven and remains rigid in the face of internal fields.
If one then is examining dielectric effects that result from the
polarization of the liquid crystal in the more conventional sense,
then perhaps the uniaxial assumption is not so strange as may seem
at first sight.  In this event, one can characterize
$\ten{\epsilon}$ by $\epsilon_{\parallel}$ along $\n$ and
$\epsilon_{\perp}$ in the directions perpendicular to $\n$.  In this
frame, one can write $\ten{\epsilon}$ simply and then extract the
element required for (\ref{eq:appenergy}):
 \bea
\ten{\epsilon}^{-1} &=& \left(  \frac{1}{\epsilon_{\parallel}} -
\frac{1}{\epsilon_{\perp}}\right)\n\,\n + \frac{1}{\epsilon_{\perp}}
\ten{\delta}\label{eq:app:tensor} \\
\left( \ten{\epsilon}^{-1}\right)_{s's'} &=& \frac{1}{\epsilon_{\perp}} +\left(
\frac{1}{\epsilon_{\parallel}} - \frac{1}{\epsilon_{\perp}}\right) \left(\n\cdot \vec{s'} \right)^2
\label{eq:app:element}
\eea
with $\n\cdot \vec{s'} = \cos\theta \cos\chi + \sin\theta \sin\chi
\cos\phi$. Rearrangement with the aid of eqn~(\ref{eq:anglechi}) gives
$\n\cdot \vec{s'} = \cos\theta \sec\chi$.  This expression and further
use of eqn~(\ref{eq:anglechi}) reduces (\ref{eq:app:element}) to:
 \be
\left( \ten{\epsilon}^{-1}\right)_{s's'} = \frac{1}{\epsilon_{\perp}} +\left(
\frac{1}{\epsilon_{\parallel}} - \frac{1}{\epsilon_{\perp}}\right) (\cos^2\theta +
\sin^2\theta\cos^2\phi)
 \label{eq:app:explicit} .
\ee

Thus the element denoted by $1/\epsilon$ in our energy expressions
has some $\phi$-dependence, but it is weak compared with the
dependence we concentrate on in the energy.

\end{document}